# Structure and frictional properties of Langmuir-Blodgett films of Cu nanoparticles modified by dialkyldithiophosphate

Jun Xu, Shuxi Dai, Gang Cheng, Xiaohong Jiang,
Xiaojun Tao, Pingyu Zhang , Zuliang Du

*Key Laboratory for Special Functional Materials, Henan University, Kaifeng 475001, China*

\* Corresponding authors. Tel.: +86 378 2867282; fax: +86 378 2867282.   zld@henu.edu.cn

**Abstract**

Langmuir-Blodgett (LB) films of dialkyldithiophosphate (DDP) modified Cu nanoparticles were prepared. The structure, microfrictional behaviors and adhesion of the LB films were investigated by transmission electron microscopy (TEM), X-ray photoelectron spectroscopy (XPS) and atomic/friction force microscopy (AFM/FFM). Our results showed that the modified Cu nanoparticles have a typical core–shell structure and fine film-forming ability. The images of AFM/FFM showed that LB films of modified Cu nanoparticles were composed of many nanoparticles arranged closely and orderly and the nanoparticles had favorable behaviors of lower friction. The friction loop of the films indicated that the friction force was affected prominently by the surface slope of the Cu nanoparticles and the microfrictional behaviors showed obvious ''ratchet effect''. The adhesion experiment showed that the modified Cu nanoparticle had a very small adhesive force.

## 1. Introduction

Interfacial friction is one of the oldest problems in physics and chemistry. Due to its practical importance and the relevance to basic scientific questions, there has been a major increase in the activity of the study of interfacial friction on the microscopic level recently [1].

The progress of high technologies such as micro/nanoelectromechanical systems (MEMS/NEMS) brought forward an austere challenge on the traditional tribology, which urged people to investigate friction phenomena in nanoscale and develop nanolubrication technologies. The invention of the atomic and friction force microscope (AFM/FFM) [2,3] has provided an ideal solution for employing micro/nanotribological techniques to study the friction and wear processes of micro- and nano-structured thin films, as demonstrated by many researchers [4–6].

It is well known that LB films are ideal to model lubricant films for fundamental studies of tribological property, because they can form densely packed and ordered structures on solid surfaces [7–10]. In order to improve the antifriction performance of LB films, in our former work, the tribological behaviors of LB films of the metal compound modified by DDP have been studied [11–14]. While metal nanoparticles are a kind of lubrication additives with fine performance, which has been reported on many papers [15–19], spherical metal nanoparticles can act as ''ball-bearings'' between the frictional counterparts, so they can efficiently improve tribological properties of lubricant. In this paper, the DDP modified Cu nanoparticles LB films were prepared and their characteristics of structure, friction and adhesion were investigated. The results are helpful in improving the understanding of the friction and wear mechanisms of ultra thin ordered molecular lubricating films.

## 2. Experimental

### 2.1. Materials

The DDP modified Cu nanoparticles were prepared by an organic synthesis method. A typical synthesis procedure has been described in our previous work [20–23]. The modified Cu nanoparticles exhibited good dispersivity in organic solvents such as acetone, toluene, and chloroform.

\* Corresponding authors. Tel.: +86 378 2867282; fax: +86 378 2867282.
  *E-mail address:* zld@henu.edu.cn (Z. Du).

### 2.2. LB film preparation

Surface pressure–molecular area ($\pi$–$A$) isotherm of DDP modified Cu nanoparticles was obtained using a computer-controlled Langmuir trough (LB 105 ATMETA). Solutions of the modified Cu nanoparticles in chloroform were spread uniformly over a subphase of water at 21 °C. At least 30 min was allowed for evaporation of the solvent prior to compression. The film was compressed at a rate of 10 cm$^2$/min and the $\pi$–$A$ isotherm was recorded at the same time. The LB films of monolayer and three layers were transferred at the surface pressure of 20 mN/m on mica flake or copper grid and glass slice with a dipping speed of 3 mm/min by vertical lifting deposition mode.

### 2.3. Instruments

TEM observation was performed with a JEM-100 CX transmission electron microscopy (TEM). X-ray photoelectron spectroscopy (XPS) was used with an AXIS ULTRA. The Al K$\alpha$ radiation was used as the excitation source at a power of 150 W and pass energy of 40 eV. A commercial AFM/FFM (SPA 400, Seiko) was employed to study the structural properties, adhesion and microfrictional properties. The AFM/FFM used here can provide simultaneous measurements of surface topography and friction force. The sample is mounted on a piezoelectric tube (PZT) scanner that can precisely scan the sample in the horizontal ($x$–$y$) plane and can move the sample in the vertical ($z$) direction. All scanning was performed in a direction perpendicular to the long axis of the cantilever beam, which is the norm for friction measurements. In our experiment, the cantilever is made of Si$_3$N$_4$ and its spring constant is 0.02 N/m. A tip radius is about 20 nm. The normal force is 7.192 nN.

## 3. Results and discussion

### 3.1. Structure and morphology characterizations

The microstructure of DDP modified Cu nanoparticles was discussed in detail in our previous work [20,22] and typical TEM images are showed in Fig. 1A and B. As shown in Fig. 1B, the diameter of DDP modified Cu nanoparticle is about 20 nm, and the thickness of shell is about 2.8 nm. There are double 18 alkyl chains for DDP molecule, and the length of stearic acid with 18 carbon chains is about 2.58 nm [24], which indicates Cu nanoparticles were completely covered by DDP, and the modified nanoparticle shows a typical core–shell structure (see Fig. 1C). Fig. 1D presents the TEM image of the monolayer. We can see that this monolayer is composed of nanoparticles with diameter of 24 nm. These modified nanoparticles are uniform and compact with slight aggregation at some parts, which showed these nanoparticles did not form multilayer aggregation and this result was consistent with the followed $\pi$–$A$ isotherm.

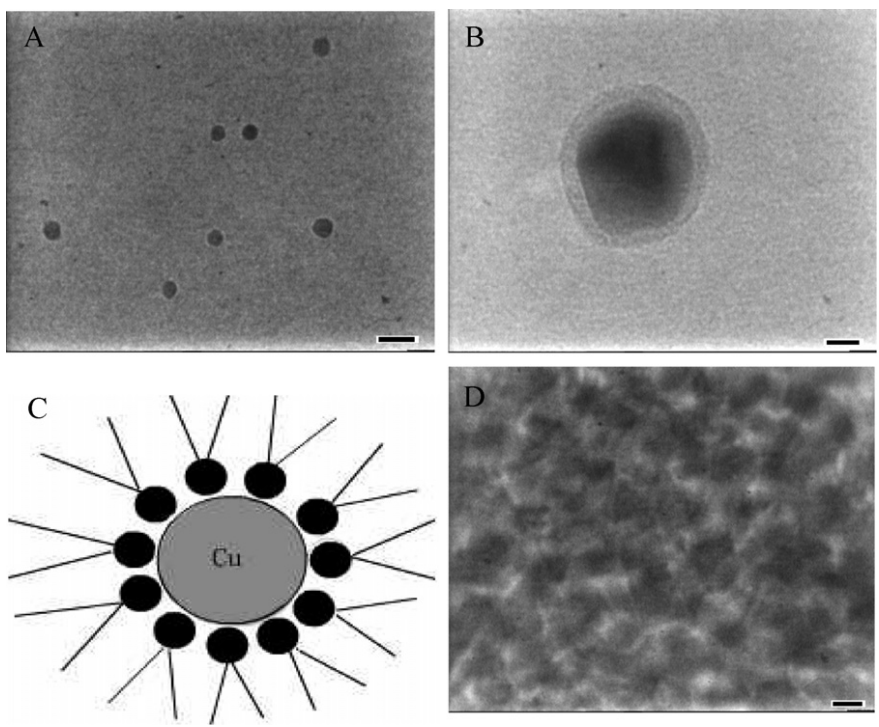

Fig. 1. TEM images of DDP modified Cu nanoparticles and the ideal model of the structure. (A and B) The DDP modified Cu nanoparticles dropped onto copper grid. (C) The ideal structure model of DDP modified Cu nanoparticles. (D) The monolayer deposited onto copper grid under 20 mN/m (A, bar = 18 nm; B, bar = 5 nm; D, bar = 6 nm).

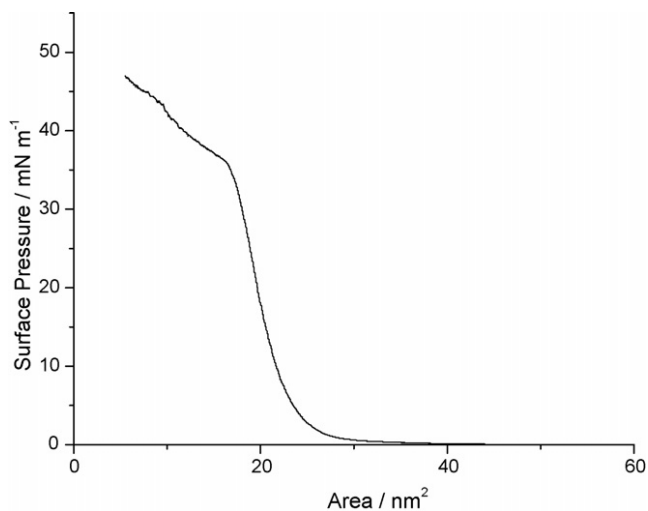

Fig. 2. π–A isotherm of DDP modified Cu nanoparticles Langmuir film at room temperature.

### 3.2. π–A isotherm

Fig. 2 shows typical surface pressure–area (π–A) isotherm of the modified Cu nanoparticles dealt with by centrifugation at room temperature. We can see that DDP modified Cu nanoparticles formed a steady Langmuir film. The surface pressure increases gradually with decreasing of area, and when it reached 37 mN/m, this monolayer begins to collapse. It is well known that although oxide or metal nanoparticles modified by organic molecules are hydrophobic, they can form steady Langmuir film at the air–water interface [25–32]. In this case, the nanoparticles modified with organic molecules have low surface tension and can be diffused to form steady Langmuir monolayer on the water surface.

It is clear that the collapse pressure of DDP modified Cu nanoparticles monolayer is lower than that of pure DDP monolayer, because DDP is a kind of typical amphiphilic molecule and has good film-forming capability and its collapse pressure is about 64.7 mN/m. In fact, there are usually free DDP molecules among DDP modified Cu nanoparticles because free DDP has strong Van der Waals force with alkyl chains of DDP modified Cu nanoparticles. In order to enhance the purity of nanoparticles, we tried our best to remove free DDP molecules through centrifugation. The collapse pressure is about 47 mN/m for DDP modified Cu nanoparticles without centrifugation process, and the collapse pressure decreases with increase of centrifugal times. After three times centrifugations, the collapse pressure reaches 37 mN/m and keeps unchanged, just as shown in Fig. 2. It indicated that the free DDP molecules were almost removed completely from the solution, which would offer a guarantee for latter study of microfrictional behaviors.

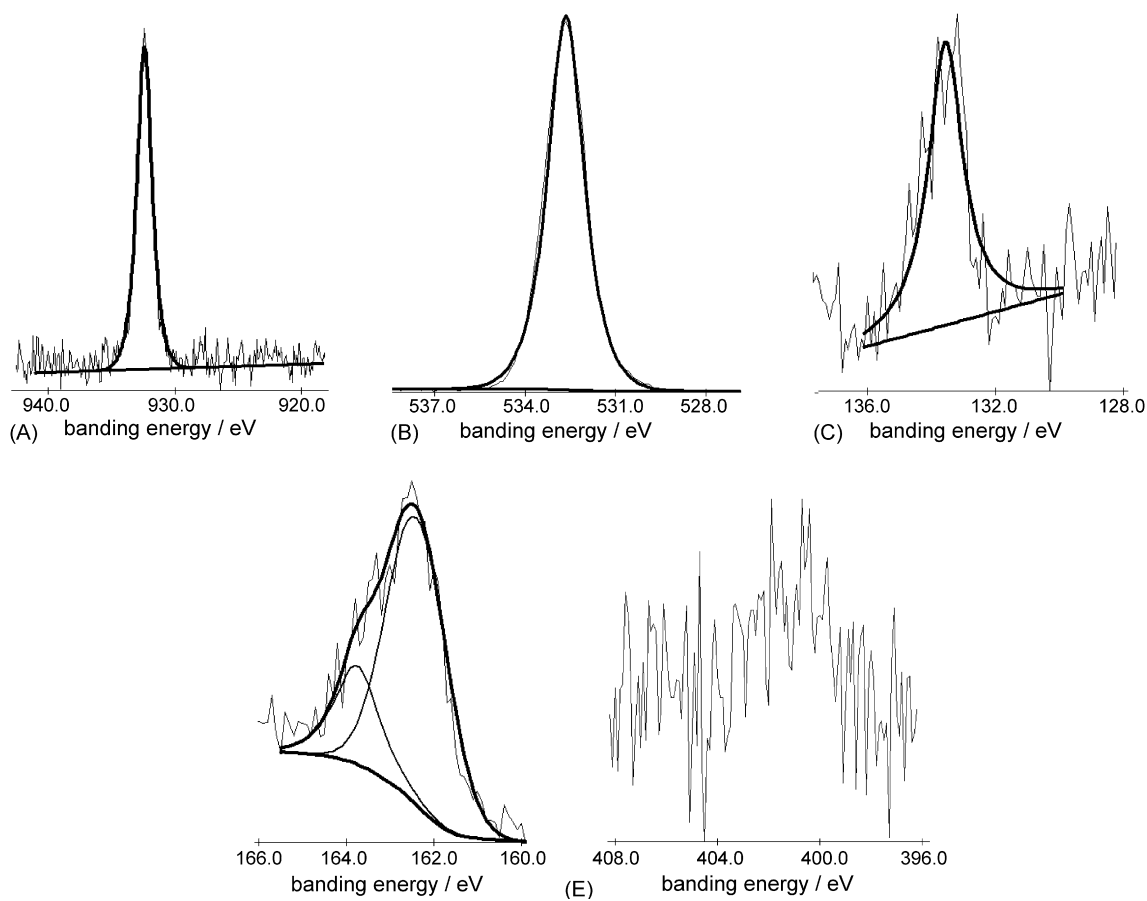

Fig. 3. The XPS spectra of three layers LB films of DDP modified Cu nanoparticles on the glass under 20 mN/m. (A) Cu $2p_{3/2}$; (B) O $2p_{3/2}$; (C) P $2p_{3/2}$; (D) S $2p_{3/2}$; (E) N.

## 3.3. XPS analysis

Fig. 3 shows the XPS spectra of three layers of LB films of DDP modified Cu nanoparticles on the glass under the pressure of 20 mN/m. The binding energy 284.8 eV of polluted carbon is used as the criterion. We can see that binding energies of Cu, O, P, S correspond to 932.4, 532.6, 133.6, 162.4 and 163.8 eV, respectively. Compared with standard spectrum, the valence of Cu is between 0 and +1, namely, Cu is bonded to S and P of PyDDP, which also indicates Cu nanoparticles were not oxidized. This result showed the surface of Cu nanoparticle was covered by DDP completely, which agreed with our previous work [18,20]. In addition, Fig. 3E shows that the signal of N in LB film of modified Cu is absent, which indicates that there is not or little free DDP in the LB film, which is consistent with the result of $\pi$–A isotherm.

## 3.4. AFM/FFM measurements

Fig. 4A and B show the topography and frictional image of LB monolayer of DDP modified Cu nanoparticles (transfer pressure is 20 mN/m, mica substrate) acquired simultaneously at the same sample location. As shown in Fig. 4A, the LB film of DDP modified Cu nanoparticles is composed of many small brighter dots, and these dots correspond to the modified Cu nanoparticles. Besides, this film is smooth and compact in the whole, with the maximum protuberance about 100 nm, and the sizes of modified Cu nanoparticles are average 50–100 nm. In comparison, the topography (A) and the friction image (B) correspond well; besides, the brighter nanoparticles in the topographies correspond to the dark areas in the friction images, while dark regions in the friction images indicate lower friction which show that DDP modified Cu nanoparticle has the function of decreasing friction.

We consider that the reasons of the nanoparticles in LB films having the function of lowering friction lie in two aspects: one is the special core–shell structure of nanoparticles, there is a monolayer of DDP molecules around the Cu core, this kind of modificatory material includes double 18 fatty chains and the interactional force between the chains is Van der Waals force, which indicates these alkyl chains bend easily under the operation of small shearing force and that results in the small shearing resistance [33]; the other is that these nanoparticles with special structure can also decrease friction with rolling effect easily during the friction [34].

Comparing Fig. 4A and B carefully, we find that the shapes of dark regions in friction image are not consistent with the circular brighter dots in topography completely and compressed to elliptical shapes in landscape orientation. The local profile images can show us this detail distinctly. Fig. 4C and D are profile images of the corresponding selected sections in A and B, respectively. Comparing 4C with 4D, we can find that the highest locations in topography do not correspond to the lowest locations in friction image completely and the highest locations in friction image do not correspond to the lowest locations in topography similarly, they stagger about half slope each other, which shows obvious character of topography modulation.

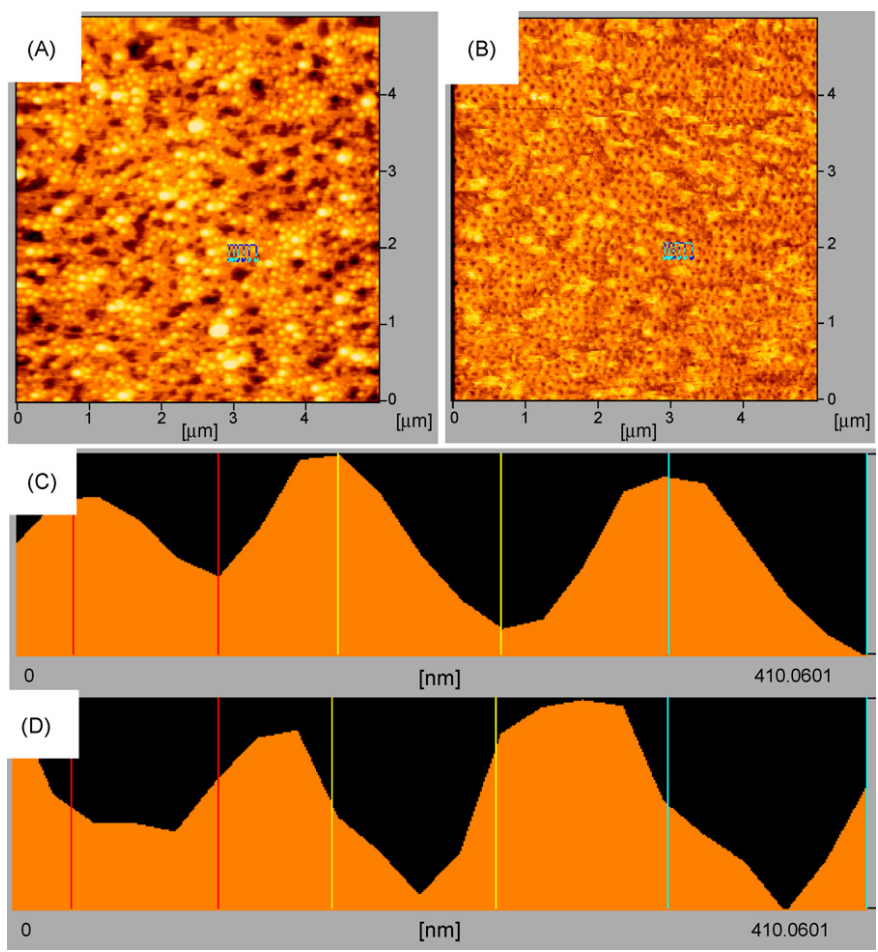

Fig. 4. AFM images of DDP modified Cu nanoparticles monolayer LB film on mica. (A) Topography; (B) friction image; (C and D) profile images corresponding to the points of topography and friction images. (A) and (B) were taken at an applied force of 7.196 nN. Dark regions in the friction images indicate lower friction.

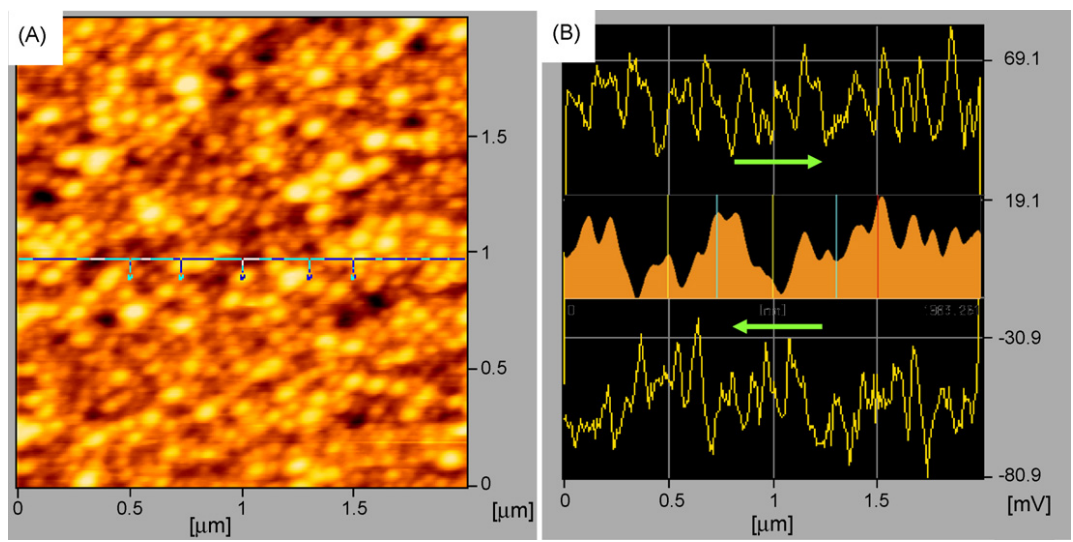

Fig. 5. Frictional curve and profile images corresponding to topography image of DDP modified Cu nanoparticles monolayer LB film on mica. The normal force is 7.196 nN. (A) Topography image; (B) frictional curve and profile images corresponding to topography image. Scan area is 2 μm × 2 μm.

In order to investigate the influence of topography-induced effects on the friction force distinctly, we investigated a typical friction loop with the change of surface topography. Fig. 5A shows the topography, Fig. 5B is a representative plot of the frictional forces in a voltage signal versus the lateral displacement for the LB film. The upper and lower curves indicate the trace and retrace friction profiles, respectively. From Fig. 5B, we can see the change of frictional force corresponds to that of surface topography, which indicates friction may be induced by topography [35]. When we observe friction loop and profile image corresponding to topography of the LB film of DDP modified Cu nanoparticles monolayer carefully, we can find these regions with nanoparticles such as 500 and 1500 nm correspond to lower friction, while the areas without nanoparticles such as 1000 nm have the higher frictional signal, which also reveal DDP modified Cu nanoparticles have the function of decreasing friction. In addition, it is clear that the peaks of friction do not correspond to that of topography but to the gradient variations. Namely, friction force increases suddenly when tip scans the peak of asperities and decreases when tip leaves the asperities suddenly.

It is well known that the measured friction forces (or lateral forces) are generated by both material effects as well as topography-induced effects when an AFM tip is scanned across a sample surface [36]. Ruan and Bhushan [37] have found the better corresponding relation between the image of friction and the variety of gradient when they investigated frictional characteristic of high directional pyrogenation plumbaginous and put forward ratchet mechanism. When a tip slides over the leading (ascending) edge of an asperity, the slope is positive, while it is negative during sliding over the trailing (descending) edge of the asperity. Thus, friction is high at the leading edge of asperities and low at the trailing edge. Since we noticed the applied normal force is small, and damage to the sample surface is also small, the contribution of ploughing is expected to be negligible. The monolayer LB film of DDP modified Cu nanoparticles is uniform, the diameters of these nanoparticles are 50–100 nm, so it is not ignored that the surface undulation influences on the friction signal. What's more, the ratchet mechanism is operative in our experiment because the probe has a tip of radius about 20 nm, while the diameters of modified Cu nanoparticles are between 50 and 100 nm. Namely, the ratchet mechanism should be dominant for the local variations in the friction profiles.

From above analysis, we can understand easily the phenomenon that the topography does not correspond to the image of friction completely. When the tip scans the sample, friction force will begin to reduce because those nanoparticles have a function of decreasing friction, but the torsion of cantilever which is influenced by slope increases rapidly, which results in the highest value of friction force at the ascending half slope; and when tip reaches the highest locality of asperity, the influence of slope on cantilever disappears, friction force shows lower value, in this time, the friction force is the real value of nanoparticle and that indicates real frictional behavior of modified nanoparticle; when tip leaves nanoparticle, the torsion of cantilever influenced by slope should be decreased, friction force reaches the lowest value at the descending half slope. When tip leaves the nanoparticle completely, friction force begins to increase, when tip meets next nanoparticle, friction force reaches the highest value again under the influence of slope, and above process repeats again.

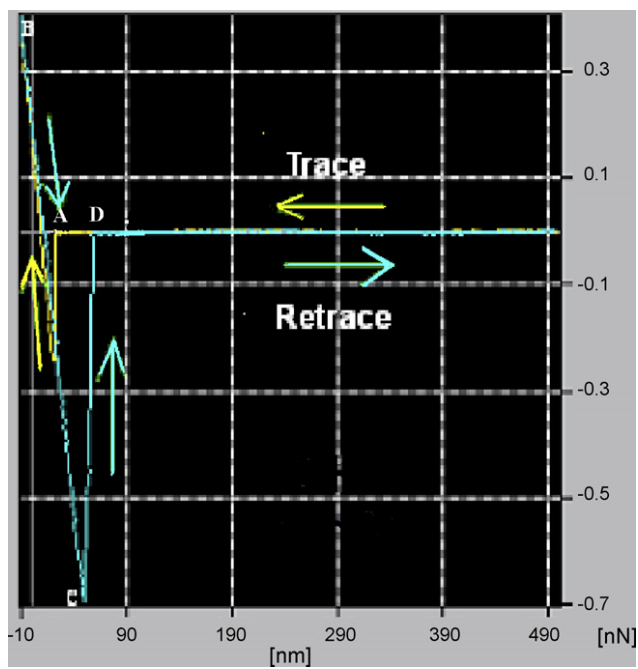

Fig. 6. Force–distance curve measured on a DDP modified Cu nanoparticle. Contact occurs at A, retraction begins at C, the tip breaks free of the adhesive forces of the sample at D and returns to free air.

## 3.5. Adhesive force measurement

The force–distance curve recorded from DDP modified Cu nanoparticle is shown in Fig. 6. The horizontal axis gives the distance the piezo travels while the vertical axis gives the magnitude of adhesive force. The piezo is constrained to move vertically. Point A is called the "pull on" point, at which the cantilever suddenly springs in contact with the sample due to short-range attractive forces. From this point on, the tip is in contact with surface, and the piezo extends until the cantilever attains the preset trigger deflection value B, then it begins to retract. At point C, the tip goes beyond the zero deflection (flat) line, because of the attractive forces (Van der Waals forces and long-range meniscus forces), into the adhesive force regime. The cantilever elastic force equals the adhesive force and just after this point, the cantilever snaps back to free state at point D. This point is also called the "pull off" point, at which the cantilever pulls off from the sample and adhesive force is then called the "pull off" force. From Fig. 6, the adhesive force of about 0.7 nN can be obtained, which is far below than that of distearoylphosphatidylethanolamine (DSPE) and dioleoylphosphatidylethanolamine (DOPE) with identical hydrophilic head groups that had been found to be 10.5 and 6.1 nN, respectively [38]. It indicates that DDP modified Cu nanoparticles have good ability of adhesion-resistance.

## 4. Conclusion

We investigated the structure, frictional properties and adhesion of LB films of DDP modified Cu nanoparticles. The results showed that the modified Cu nanoparticle has typical core–shell structure and good film-forming capability. The LB film was composed of many nanoparticles arranged closely and orderly. The sizes of these nanoparticles in monolayer film are 50–100 nm. These nanoparticles in film have the function of lowering friction force and the microscale friction varies with local slope of sample surface. Ratchet mechanism is operative in this scale. In addition, the film has very small adhesive force of about 0.7 nN, which implies it has fine performance of adhesion-resistance.

## Acknowledgements

This work was supported by Natural Science Foundation of China (no. 90306010, 20371015, 20571024) and Program for New Century Excellent Talents in University of China.